\documentclass[epj]{svjour}

\usepackage{graphicx}
\usepackage{fancyhdr}
\usepackage{amsmath,color,axodraw}

\def\mytitle{My title} 
\def\myauthors{My name}  
\def\mytype{My type of session}
\def\mysession{My session}

\def\mytitle{Tau polarization in SUSY cascade decays at LHC}
\def\myauthors{Kentarou Mawatari}
\def\mytype{Contributed Talk}    
\def\mysession{Colliders - SUSY Phenomenology}

\newcommand{\nn}{\nonumber}
\newcommand{\sq}{\tilde q}
\newcommand{\sn}{\tilde\chi^0_2}

\newcommand{\st}{\tilde\tau}
\newcommand{\fn}{\tilde\chi^0_1}

\begin{document}
\title{Tau polarization in SUSY cascade decays at LHC%
  \thanks{presented at the 15th International Conference on
  Supersymmetry and the Unification of Fundamental Interactions
  (SUSY07), July 26 - August 1, 2007, Karlsruhe, Germany.}}
\author{Kentarou Mawatari
\thanks{\emph{Email:} kentarou@kias.re.kr}%
}                     
%
%
\institute{School of Physics, Korea Institute for Advanced Study (KIAS),
           Seoul 130-722, South Korea
          }
%
\date{}
\abstract{
 We explore how the polarization of the tau leptons in the cascade decay
 $\sq\to q\sn\to q\tau\st\to q\tau\tau\fn$ can be exploited to study
 mixing properties of neutralinos and staus. 
 We present details of the analysis including experimental effects such as
 transverse momentum cuts for the $\tau\to\pi\nu$ decay
 mode, and show that the di-pion invariant mass
 distribution provides valuable information on their properties.
\PACS{
      {14.80.Ly}{Supersymmetric partners of known particles} \and
      {13.85.Hd}{Inelastic scattering: many-particles final states}
     } 
} 
\maketitle
%

\section{Introduction}

At LHC squarks and gluinos will be copiously produced, and
cascade down, generally in several steps, to the lightest supersymmetric
particle (LSP), which is stable in $R$-parity conserving scenarios. 
Though the LSPs escape detection, the kinematic edge measurements 
of the visible final
state in various combinations of quark jets and leptons can serve to
study the precision with which the masses of supersymmetric particles
can be measured at LHC, for a summary see Ref.~\cite{Weiglein:2004hn}.
In addition, invariant mass distributions have been studied to determine
 the slepton mixing~\cite{Goto:2004cpa} and the spin of the particles
 involved~\cite{Barr:2004ze,Smillie:2005ar,Datta:2005zs,Wang:2006hk}, 
 shedding light on the nature of the new
particles observed in the cascade and on the underlying physics scenario.

So far, cascades have primarily been studied involving first and second
generation leptons/sleptons. Recently we explored how the
polarization of $\tau$ leptons can be exploited to study $R/L$ chirality
and mixing effects in both the neutralino and the stau 
sectors~\cite{Choi:2006mt}.
In this report, we demonstrate whether measuring the correlation of the
$\tau$ polarizations provides an excellent instrument to analyze these
effects,%
\footnote{For a discussion of polarization effects in single $\tau$
decays see Ref.~\cite{Nojiri:1994it}.}%
using a event generator.

\section{Tau polarization analyzer}

As polarization analyzer we use single pion decays of the $\tau$'s.
At high energies the mass of the $\tau$ leptons can be neglected and the
fragmentation functions are linear in the fraction $z$ of
the energy transferred from the polarized $\tau$'s to the
$\pi$'s~\cite{Bullock:1992yt}:
\begin{subequations}
\begin{align}
 (\tau_R)^\pm &\to\ \stackrel{(-)}{\nu_\tau}\pi^\pm : && F_R=2z,
 \label{eq:tau_to_pi_R}\\
 (\tau_L)^\pm &\to\ \stackrel{(-)}{\nu_\tau}\pi^\pm : && F_L=2(1-z).
 \label{eq:tau_to_pi_L}
\end{align}
\end{subequations}
In the relativistic limit, helicity and chirality are of equal and opposite
sign for $\tau^-$ leptons and $\tau^+$ anti-leptons, respectively. 
For notational convenience we characterize the $\tau$ states by chirality.

This report should serve only as an exploratory theoretical study. Experimental
simulations will include other $\tau$ decay final states in addition to $\pi$'s,
{\it e.g.} $\rho$'s and $a_1$'s. The $\rho$-meson mode is expected to contribute to
the $\tau$-spin correlation measurement even if the $\pi^\pm$ and $\pi^0$ energies
are not measured separately. In this case the $\tau$ polarization analysis power
of the $\rho$ channel is $\kappa_\rho
=(m^2_\tau-2m^2_\rho)/(m^2_\tau+2 m^2_\rho)\sim 1/2$ in contrast to $\kappa_\pi=1$,
but its larger branching fraction of {\rm B}$_\rho\approx 0.25$, {\it vs.}
{\rm B}$_\pi \approx 0.11$, more than compensates for the reduced analysis power.
Moreover, in actual experiments it should be possible to measure the $\pi^\pm$
energy and the $\gamma$ energies of the $\pi^0$'s, all emitted along the parent
$\tau$-momentum direction at high energies. Significant improvement of the $\tau$
analysis power is therefore expected from the $\rho$ and $a_1$ modes by determining
the momentum fraction of $\pi^\pm$ in the collinear limit
of their decays~\cite{Bullock:1992yt}.
For each mode, cuts and efficiencies for $\tau$ identification must be included
to arrive finally at realistic error estimates. The large size of the polarization
effects predicted on the theoretical basis, and exemplified quantitatively by the
pion channel, should guarantee their survival in realistic experimental environments,
and we expect that they can be exploited experimentally in practice.

\section{Squark cascade decay}

We consider the squark cascade decay involving third generation
leptons/sleptons 
\begin{align}
 \sq_L\to q\sn\to q\tau^{\pm}\st^{\mp}\to q\tau^{\pm}\tau^{\mp}\fn.
\label{cascade}
\end{align}
At the SPS1a point~\cite{Allanach:2002nj}, $\sn$ and $\fn$ are wino- and
bino-like, respectively. Therefore left-handed squarks decay to $\sn$
with a branching ratio $\sim 30\%$, while right-handed squarks directly
decay to the LSP $\fn$.

\begin{figure}
{\color{black}
\begin{center}
\begin{minipage}{0.45\textwidth} 
\begin{picture}(180,105)(-10,150)
\DashLine(0,170)(50,170){2}%
\Text(50,170)[c]{\large $\bullet$}
\Text(25,155)[c]{\color{red} $\tilde{q}_{\alpha}^p$}
\Text(45,200)[c]{$q_{\alpha}^p$}
\Line(50,170)(60,210)
\Line(50,170)(100,170)
\Text(100,170)[c]{\large $\bullet$}
\Text(75,155)[c]{\color{red} $\tilde{\chi}^0_j$}
\Text(95,200)[c]{$\tau^{a}_\beta$}
\Text(105,240)[c]{$\pi^{a}$}
\Line(100,170)(110,210)
\Text(110,210)[c]{\large $\bullet$}
\DashLine(110,210)(120,250){2}
\DashLine(100,170)(150,170){2}%
\Text(125,155)[c]{\color{red} $\tilde{\tau}^{-a}_{k}$}
\Text(145,200)[c]{$\tau^{-a}_\gamma$}
\Text(155,240)[c]{$\pi^{-a}$}
\Line(150,170)(160,210)
\Text(150,170)[c]{\large $\bullet$}
\Text(160,210)[c]{\large $\bullet$}
\DashLine(160,210)(170,250){2}
\Line(150,170)(200,170)
\Text(175,155)[c]{\color{red} $\tilde{\chi}^0_1$}
\end{picture}
\end{minipage}
\caption{The general structure of the quantum numbers of the particles
        involved in the squark cascade (\ref{cascade}).}
\label{fig:cascade}
\end{center}
}
\end{figure}
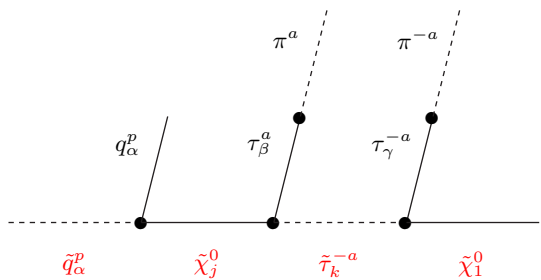

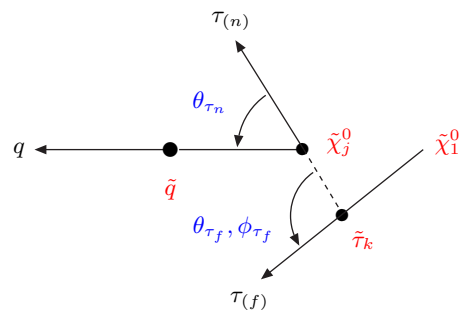
\begin{figure}
{\color{black}
\begin{center}
\begin{minipage}{0.45\textwidth} 
\begin{picture}(180,120)(-10,-10)
\Text(15,50)[r]{$q$}
\LongArrow(70,50)(20,50)
\Line(73,50)(120,50)
\Text(70,50)[c]{\Large $\bullet$}
\Text(120,50)[c]{\large $\bullet$}
\Text(85,68)[c]{\color{blue} $\theta_{\tau_n}$}
\LongArrowArc(120,50)(25,125,178)
\LongArrow(120,50)(95,90)
\Text(93,99)[]{$\tau_{(n)}$}
\DashLine(120,50)(135,25){2}
\Text(135,25)[c]{\large $\bullet$}
\Text(70,35)[c]{\color{red} $\tilde{q}$}
\Text(134,53)[c]{\color{red} $\tilde{\chi}^0_j$}
\Text(143,16)[c]{\color{red} $\tilde{\tau}_k$}
\Text(94,20)[c]{\color{blue} $\theta_{\tau_f}, \phi_{\tau_f}$}
\LongArrowArc(135,25)(20,125,214)
\Line(135,26)(165,50)
\Text(170,53)[l]{\color{red} $\tilde{\chi}^0_1$}
\LongArrow(135,26)(105,2)
\Text(100,-7)[c]{$\tau_{(f)}$}
\end{picture}
\end{minipage}
\caption{The angular configuration of the particles in the
        $\sq/\tilde{\chi}^0_j/\tilde{\tau}_k$ rest frames.}
\label{fig:frame}
\end{center}
}
\end{figure}

The distribution of the visible final state particles in the squark
cascade can be cast, for massless quark and no squark mixing,
in the general form~\cite{Goto:2004cpa}:
\begin{align}
 \frac{1}{\Gamma_{\tilde{q}_\alpha}}&\frac{d\Gamma^{pa;jk}_{\alpha\beta\gamma}}{
  d\cos\theta_{\tau_n}d\cos\theta_{\tau_f}d\phi_{\tau_f}} \nn\\
  &= \frac{1}{8\pi}
      {\rm B}(\tilde{q}_\alpha\to q_\alpha\tilde{\chi}^0_j)\,
      {\rm B}(\tilde{\chi}^0_j\to \tau_\beta \tilde{\tau}_k)\,
      {\rm B}(\tilde{\tau}_k\to\tau_\gamma\tilde{\chi}^0_1) \nn\\
  &\quad \times\left[ 1+(pa)(\alpha\beta) \cos\theta_{\tau_n} \right]
\label{eq:susy_distribution}
\end{align}
with $p=+$, $\alpha=-$, $j=2$ and $k=1$ for the squark chain of Eq.~(\ref{cascade}).
The structure of the quantum numbers in the cascade is depicted in
Fig.~\ref{fig:cascade}
while the configuration of the particles in the 
$\sq$/$\tilde{\chi}^0_2$/$\tilde{\tau}_1$
rest frames
is shown in Fig.~\ref{fig:frame}. For clarity the definitions
are summarized in the following table:
\begin{align*}
 p      &= \pm :   && \mbox{particle/anti-particle,} \\
 a      &= \pm :   && \mbox{$\tau$ and $\pi$ charge,} \\
 j      &= 2,3,4 : && \mbox{neutralino mass index,} \\
 k      &= 1,2 :   && \mbox{$\tilde{\tau}$ mass index,} \\
 \alpha &= \pm :   && \mbox{$\tilde{q}$ and $q$ $R/L$ chirality,} \\
 \beta  &= \pm :   && \mbox{near $\tau_{n}$ $R/L$ chirality,} \\
 \gamma &= \pm :   && \mbox{far $\tau_{f}$ $R/L$ chirality.} 
\end{align*}
Near $(n)$ and far $(f)$ indices denote $\tau$ and $\pi$ particles
emitted in $\tilde{\chi}^0_j$ and $\tilde{\tau}_k$ decays, respectively.

The $q_\alpha\tilde{q}_\alpha\tilde{\chi}^0_j$
and the $\tau_\beta\tilde{\tau}_k\tilde{\chi}^0_j$
vertices are given by the proper current couplings and the neutralino and
stau mixing matrix elements,
\begin{align}
 \langle \tilde{\chi}^0_j| \tilde{q}_\alpha | q_\alpha\rangle
    &= i g\, A^q_{\alpha \alpha j}, \\
 \langle \tilde{\chi}^0_j| \tilde{\tau}_k | \tau_\beta \rangle
    &= i g\, A^\tau_{\beta kj}, \ \ [\gamma \;\; {\rm correspondingly}]
\label{eq:interaction_vertex}
\end{align}
with the explicit form of the couplings $A^q_{\alpha\alpha j}$
\begin{subequations}
\begin{align}
 A^q_{LLj} &= -\sqrt{2} [T_3^q N^*_{j2}+(e_q-T^q_3) N^*_{j1} t_W], \\
 A^q_{RRj} &= +\sqrt{2} e_q N_{j1} t_W,
\end{align}
\end{subequations}
and $A^\tau_{\beta k j}$
\begin{subequations}
\begin{align}
 A^\tau_{L k j} &= -h_\tau N^*_{j3} U_{\tilde{\tau}_{k2}}
                 +\frac{1}{\sqrt{2}} (N^*_{j2}+N^*_{j1} t_W)
                  U_{\tilde{\tau}_{k1}},
\label{eq:tau_L_coupling}\\
 A^\tau_{R k j} &= -h_\tau N_{j3} U_{\tilde{\tau}_{k1}}
                -\sqrt{2} N_{j1} t_W U_{\tilde{\tau}_{k2}},
\label{eq:tau_R_coupling}
\end{align}
\end{subequations}
in terms of the $4\times 4$ neutralino mixing matrix $N$ in the standard
gaugino/higgsino basis and the $2\times 2$ stau mixing
matrix $U_{\tilde{\tau}}$ in the $L/R$ basis.
Here, $T^q_3=\pm 1/2$ and $e_q=2/3,-1/3$ are the SU(2) doublet quark isospin
and electric charge, $t_W=\tan\theta_W$ and $h_\tau=m_\tau/\sqrt{2}m_W\cos\beta$.
The distribution (\ref{eq:susy_distribution}) depends only
on the ``near
$\tau$'' angle $\theta_{\tau_n}$; this is a consequence of the scalar character
of the intermediate stau state that erases all angular correlations.

The angles in the cascade Fig.~\ref{fig:frame} are related
to the invariant masses \cite{Weiglein:2004hn,Goto:2004cpa,Smillie:2005ar},
\begin{subequations}
\begin{align}
 m^2_{\tau\tau}\!\!  &= (m^{\rm max}_{\tau\tau})^2\!\cdot
     \frac{1}{2}\left(1-\cos\theta_{\tau_f}\right), \\
 m^2_{q\tau_{n}}\!\! &= (m^{\rm max}_{q\tau_{n}})^2\!\cdot
     \frac{1}{2}\left(1-\cos\theta_{\tau_n}\right), \\
 m^2_{q\tau_{f}}\!\! &= (m^{\rm max}_{q\tau_{f}})^2\!\cdot 
     \Big[\frac{1}{4}(1+c_n)(1-c_f)
   -\frac{r_{jk}}{2} s_ns_f \cos\phi_{\tau_f} \nn \\
    &\qquad\qquad +\frac{r^2_{jk}}{4} (1-c_n)(1+c_f)\Big], 
\end{align}
\end{subequations}
where the maximum values of the invariant masses
\begin{subequations}
\begin{align}
 (m^{\rm max}_{\tau\tau})^2  &= m^2_{\tilde{\chi}^0_j}
      (1-m^2_{\tilde{\tau}_k}/m^2_{\tilde{\chi}^0_j})
      (1-m^2_{\tilde{\chi}^0_1}/m^2_{\tilde{\tau}_k}), 
  \label{ditaumax}\\
 (m^{\rm max}_{q\tau_{n}})^2 &=  m^2_{\tilde{q}_\alpha}
      (1-m^2_{\tilde{\chi}^0_j}/m^2_{\tilde{q}_\alpha})
      (1-m^2_{\tilde{\tau}_k}/m^2_{\tilde{\chi}^0_j}), \\
 (m^{\rm max}_{q\tau_{f}})^2 &= m^2_{\tilde{q}_\alpha}
      (1-m^2_{\tilde{\chi}^0_j}/m^2_{\tilde{q}_\alpha})
      (1-m^2_{\tilde{\chi}^0_1}/m^2_{\tilde{\tau}_k}),
\end{align}
\end{subequations}
$r_{jk} =m_{\tilde{\tau}_k}/m_{\tilde{\chi}^0_j}$, and abbreviations
$c_n = \cos\theta_{\tau_n}, c_f=\cos\theta_{\tau_f}$ {\it etc}. are
introduced.
Note that the rescaled invariant masses, 
$\tilde m^2\equiv m^2/(m^{\rm max})^2$, are used for analysis in our original 
paper~\cite{Choi:2006mt}.

\section{Pion invariant mass distribution}

\begin{figure}
\begin{center}
 \includegraphics[width=.315\textwidth,angle=0,clip]{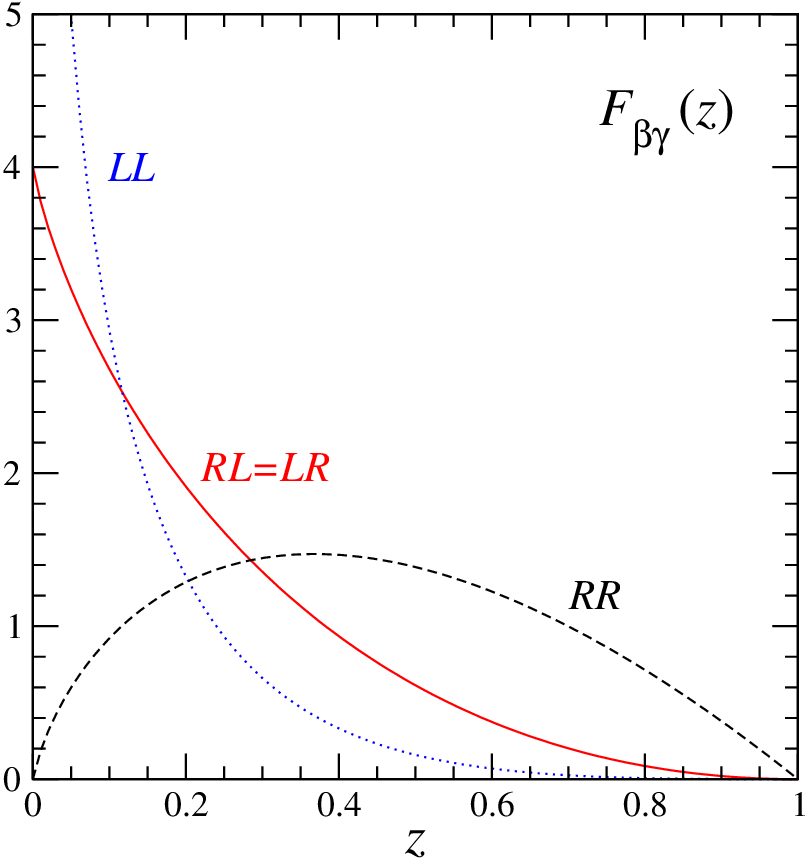}
 \caption{Double fragmentation functions $F_{\beta\gamma}(z)$.}
 \label{fig:doublef}
\end{center}
\end{figure}

Pion invariant mass distributions, summed over near and far particles,
are predicted by folding 
the original single $\tau$ and double $\tau \tau$ distributions,
$d\Gamma_\beta / dm^2_{q\tau}$
and $d\Gamma_{\beta\gamma} / dm^2_{\tau\tau}$, with the single and double
fragmentation functions $F_\beta$ and $F_{\beta\gamma}$, where the indices
$\beta,\gamma$ denote the chirality indices $R/L$ of the $\tau$ leptons.
Based on standard techniques, the following relations can be derived
for [$q \pi$] and $[\pi \pi]$ distributions:
\begin{align}
 \frac{d\Gamma}{dm^2_{q\pi}}
  &= \int^1_{m^2_{q\pi}} \frac{dm^2_{q\tau}}{m^2_{q\tau}}\,
     \frac{d\Gamma_\beta}{dm^2_{q\tau}}\,
     F_\beta\Big(\frac{m^2_{q\pi}}{m^2_{q\tau}}\Big), \\
 \frac{d\Gamma}{dm^2_{\pi\pi}}
  &= \int^1_{m^2_{\pi\pi}} \frac{dm^2_{\tau\tau}}{m^2_{\tau\tau}}\,
     \frac{d\Gamma_{\beta\gamma}}{dm^2_{\tau\tau}}\,
     F_{\beta\gamma}\Big(\frac{m^2_{\pi\pi}}{m^2_{\tau\tau}}\Big).
\end{align}
The single and double distributions,
$d\Gamma_\beta / dm^2_{q\tau}$ and
$d\Gamma_{\beta\gamma} / dm^2_{\tau\tau}$,
can be derived from Eq.$\,$(\ref{eq:susy_distribution}) by integration.
The single $\tau_{\beta} \to \pi$ fragmentation function,
{\it cf.} Eqs.$\,$(\ref{eq:tau_to_pi_R}) and (\ref{eq:tau_to_pi_L}) with
$z=m^2_{q\pi}/m^2_{q\tau}$, can be summarized as
\begin{align}
 F_\beta(z) = 1 +\beta\, (2z-1),
\end{align}
while the double $\tau_{\beta} \tau_\gamma \to \pi \pi$
fragmentation functions, with $z=m^2_{\pi\pi}/m^2_{\tau\tau}$,
are given by
\begin{subequations}
\begin{align}
 F_{RR}(z) &=  4z\log\frac{1}{z}, \\
 F_{RL}(z) &=  F_{LR}(z) = 4\Big[1-z-z\log\frac{1}{z}\Big],  \\
 F_{LL}(z) &=  4\Big[(1+z)\log\frac{1}{z}+2z-2\Big].
\end{align}
\end{subequations}
The shape of the distributions $F_{\beta\gamma}(z)$ ($\beta,\gamma=R,L$) is
presented in Fig.~\ref{fig:doublef}.
All distributions, normalized to unity, are finite except $F_{LL}$
which is logarithmically divergent for $z\to 0$.

The analysis of $\pi\pi$ invariant mass, {\it etc}., for polarization
measurements in cascades does not require the experimental
reconstruction of $\tau$-jet energies, in contrast to the fraction $R$
of charged $\pi$ over $\tau$ energy, see Ref.~\cite{Guchait:2002xh}.

\section{Simulations}

At the SPS1a point~\cite{Allanach:2002nj}, the production cross section
of the squarks ($\tilde u_L$ and $\tilde d_L$) is 33 pb at 
LHC~\cite{Weiglein:2004hn}, 
and the branching fractions of the cascade (\ref{cascade}) are
${\rm B}(\tilde{q}_L\to q\tilde{\chi}^0_2)\sim 30\%$,
${\rm B}(\tilde{\chi}^0_2\to \tau\tilde{\tau}_1)=88\%$, and 
${\rm B}(\tilde{\tau}_1\to\tau\tilde{\chi}^0_1)=100\%$. 
Hence, roughly 10$^5$ events for the cascade would be expected with
$L=10$ fb$^{-1}$. In this report, taking into account the branching fraction
${\rm B}(\tau\to\pi\nu)$, we estimate the di-$\pi$ invariant mass
distribution with 10$^3$ events. The events are generated by 
{\tt MadGraph/MadEvent}~\cite{Alwall:2007st} with the {\tt DECAY}
package for $\tau$ decays, and studied using {\tt MadAnalysis}. 

\begin{figure}
\begin{center}
 \includegraphics[width=.35\textwidth,angle=0,clip]{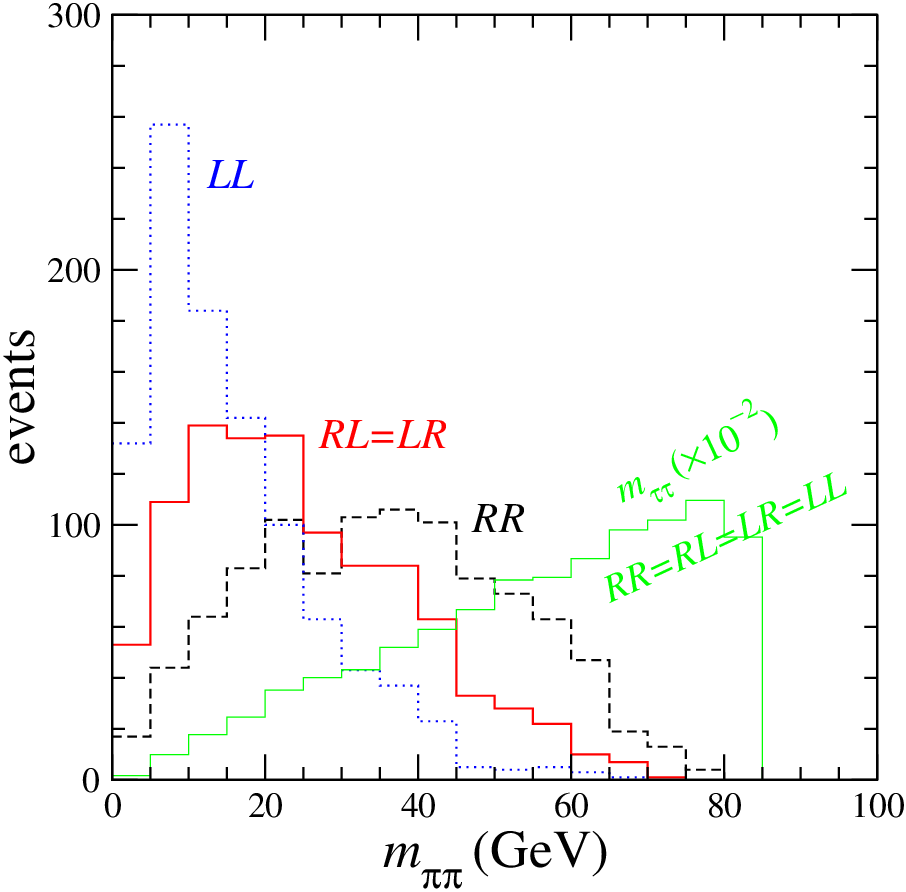}
 \caption{Di-$\pi$ invariant mass distributions in the $\sn$ decays
 of the cascade (\ref{cascade}). The indices denote the chiralities of
    the near and far $\tau$ leptons. The di-$\tau$ invariant mass
 distribution is also shown by a thin line.}
 \label{fig:x2decay}
\end{center}
\end{figure}

The great potential of polarization measurements for determining
mixing phenomena is demonstrated in Fig.~\ref{fig:x2decay},
displaying the di-$\pi$ invariant mass distributions in the
$\tilde{\chi}^0_2$ decays of the cascade 
(\ref{cascade})~\cite{Cahn:1996td,Choi:2006mt}. 
 The di-$\tau$ distribution is also shown.
The maximum value of the di-$\tau$ invariant mass in Eq.~(\ref{ditaumax})
is 84 GeV for   
the sparticle masses at SPS1a: $m_{\sq}\sim 560$ GeV, $m_{\sn}=181$
GeV, $m_{\st}=134$ GeV, and $m_{\fn}=97$ GeV.
While the lepton-lepton
invariant mass does not depend on the chirality indices of the near and
far $\tau$ leptons, the shape of the $\pi$ distribution depends
strongly on the indices, as expected.

\begin{figure}
\begin{center}
 \includegraphics[width=.35\textwidth,angle=0,clip]{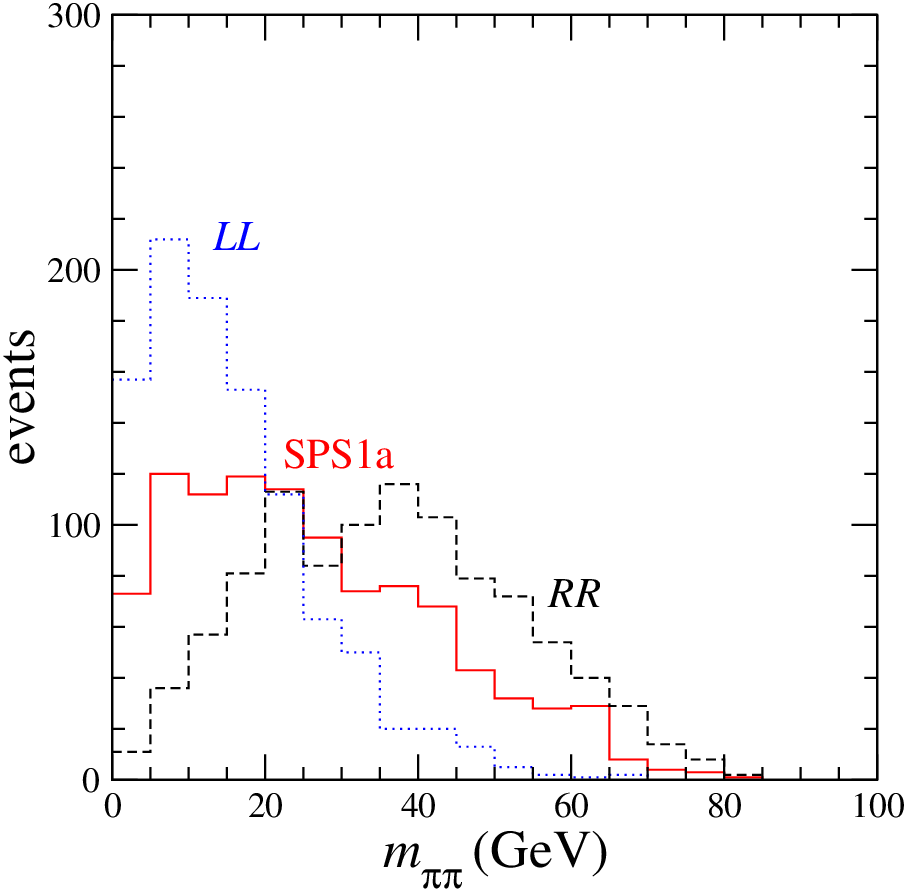}
 \caption{Di-$\pi$ invariant mass distribution in the $\sq$ decays
 of the cascade (\ref{cascade}) at SPS1a. The dashed and dotted lines
 indicate the hypothetical $RR$- and $LL$-type distributions,
 respectively, for comparison.}
 \label{fig:sqdecay}
\end{center}
\end{figure}

Fig.~\ref{fig:sqdecay} shows the di-$\pi$ invariant mass distribution in
the $\sq$ decays of the cascade (\ref{cascade}) at SPS1a. The
hypothetical $RR$- and $LL$-type distributions are also shown by
dashed and dotted lines, respectively, for comparison. 
The solid line for SPS1a clearly indicates that the near/far $\tau$
couplings are 
$RL$- or $LR$-dominated ({\it cf.} Fig.~\ref{fig:x2decay}).
This can be traced back to the fact that $\sn$ is nearly $\tilde W$-like and
$\st_1$ is nearly $\st_R$-like at SPS1a, which is reflected in the
$\tau_{\beta}\st_k\tilde\chi^0_j$ coupling, $A^{\tau}_{\beta k j}$, in 
Eqs.~(\ref{eq:tau_L_coupling}) and (\ref{eq:tau_R_coupling}).
The magnitude of the couplings are  
\begin{align}
 |A^{\tau}_{L12}| &= 0.240,  & |A^{\tau}_{R12}| &= 0.0626, \nn\\
 |A^{\tau}_{L11}| &= 0.0763, & |A^{\tau}_{R11}| &= 0.745, 
\end{align} 
and the polarization of the near (far) $\tau$ leptons, $\tau_{n(f)}$, is
given in the $m_{\tau}\ll m_{\st_1}$ limit as
\begin{align}
 P_{\tau_{n(f)}}=\frac{|A^{\tau}_{R12(1)}|^2-|A^{\tau}_{L12(1)}|^2}
                 {|A^{\tau}_{R12(1)}|^2+|A^{\tau}_{L12(1)}|^2}.
\end{align}
Therefore, the polarization of the near $\tau$ is almost left-handed 
($P_{\tau_n}=-0.873$), while the far $\tau$ has the right-handed
polarization ($P_{\tau_f}=+0.979$). This leads the $LR$-dominated
distribution in the di-$\pi$ invariant mass. 

Finally we briefly study the effects of the experimental cut on the
$\pi$ transverse momenta.
Experimental analyses of $\tau$ particles are a difficult task at LHC.
Isolation criteria of hadron and lepton tracks must be met which reduce the
efficiencies strongly for small transverse momenta.
Stringent transverse momentum cuts
increase the efficiencies but reduce the primary event number and erase
the difference 
between $R$ and $L$ distributions. On the other hand, fairly small
transverse momentum 
cuts reduce the efficiencies but do not reduce the primary event number and the
$R/L$ sensitivity of the distributions.
Optimization procedures in this context are far beyond the scope of this
report. 
Experimental details for di-$\tau$ final states for the cascades at LHC
may be studied in the recent notes~\cite{mangeol}.  

\begin{figure}
\begin{center}
 \includegraphics[width=.35\textwidth,angle=0,clip]{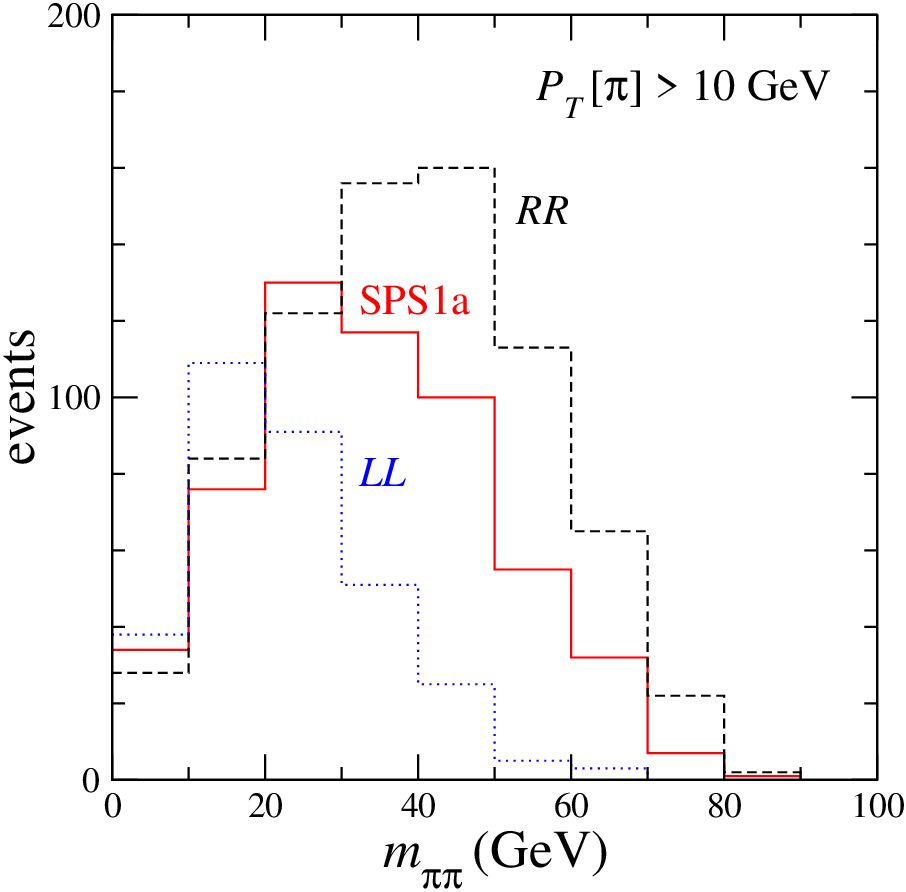}
 \caption{The same as Fig.~\ref{fig:sqdecay} with a kinematic cut of 10
 GeV on the $\pi$ transverse momenta.}
 \label{fig:sqdecay_ptcut}
\end{center}
\end{figure}

In Fig.~\ref{fig:sqdecay_ptcut} it is shown how a cut of 10 GeV on the
$\pi$ transverse momenta modifies the distributions of the
di-$\pi$ invariant mass in Fig.~\ref{fig:sqdecay}. 
75\%, 55\%, and 32\% of the events for $RR$-type, SPS1a, and $LL$-type,
respectively, survive the selection. 
The $RR$-type distribution is mildly affected while the
$LL$-type is shifted more strongly. The different size
of the shifts can be traced back to the different shapes of the $R$ and $L$
fragmentation functions. Since $L$ fragmentation is soft, more events with
low transverse momentum are removed by the cut and the shift is
correspondingly larger than for hard $R$ fragmentation.
Apparently, the peak positions of the distributions are still different,
and the transverse momentum cut does not erase the distinctive
difference between them.

\section{summary}

The analysis of $\tau$ polarization in cascade decays
provides valuable information on chirality-type and mixing of supersymmetric
particles. The exciting effects are predicted for the invariant mass
distributions in the di-$\pi$ sector generated by the two polarized $\tau$
decays. 
Note that these effects are independent of the couplings in the
$\sq/q$ sector and also of the polarization state of $\tilde{\chi}^0_2$. 
See more details in Ref.~\cite{Choi:2006mt}.

\section*{Acknowledgements}
The author thanks S.Y.~Choi, K.~Hagiwara,
Y.G.~Kim, and P.M.~Zerwas for the collaboration.

\end{document}